\documentclass[twoside,twocolumn,9pt]{article}
\usepackage{extsizes}
\usepackage[super,sort&compress,comma]{natbib} 
\usepackage[version=3]{mhchem}
\usepackage[left=1.5cm, right=1.5cm, top=1.785cm, bottom=2.0cm]{geometry}
\usepackage{balance}
\usepackage{times,mathptmx}
\usepackage{sectsty}
\usepackage{graphicx} 
\usepackage{lastpage}
\usepackage[format=plain,justification=justified,singlelinecheck=false,font={stretch=1.125,small,sf},labelfont=bf,labelsep=space]{caption}
\usepackage{float}
\usepackage{fancyhdr}
\usepackage{fnpos}
\usepackage[english]{babel}
\usepackage{array}
\usepackage{droidsans}
\usepackage{charter}
\usepackage[T1]{fontenc}
\usepackage[usenames,dvipsnames]{xcolor}
\usepackage{setspace}
\usepackage[compact]{titlesec}
\usepackage{verbatim}

\usepackage{amssymb}
\usepackage{amsmath}
\usepackage{amsfonts}
\usepackage{bm}
\usepackage{dsfont}
\usepackage{fancyhdr}
\usepackage{euscript}
\usepackage{graphicx}
\usepackage{mathrsfs}
\usepackage{textcmds}
\usepackage{color}
\usepackage{footnote}
\usepackage[pdftex,colorlinks,bookmarksopen,bookmarksnumbered,citecolor=red,urlcolor=red]{hyperref}
\usepackage{cancel}
\usepackage{textcomp}
\usepackage{algorithm} 
\usepackage[normalem]{ulem}
\usepackage[percent]{overpic}

\newcommand{\trans}{{\mskip-1mu\scriptscriptstyle\top\mskip-3mu}}
\def\tr{\mbox{tr}\,}

\graphicspath {
    {./}
    {./figures/}
}

\definecolor{cream}{RGB}{222,217,201}

\definecolor{darkpastelgreen}{rgb}{0.01, 0.75, 0.24}

\usepackage{epstopdf}

\definecolor{cream}{RGB}{222,217,201}

\begin{document}

\pagestyle{fancy}
\thispagestyle{plain}
\fancypagestyle{plain}{

\fancyhead[C]{\includegraphics[width=18.5cm]{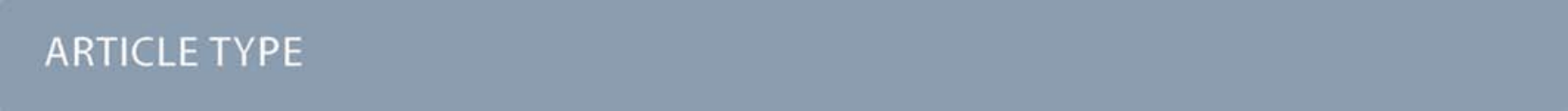}}
\fancyhead[L]{\hspace{0cm}\vspace{1.5cm}\includegraphics[height=30pt]{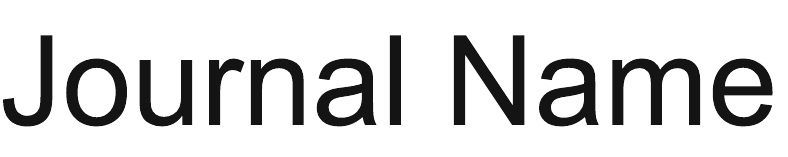}}
\fancyhead[R]{\hspace{0cm}\vspace{1.7cm}\includegraphics[height=55pt]{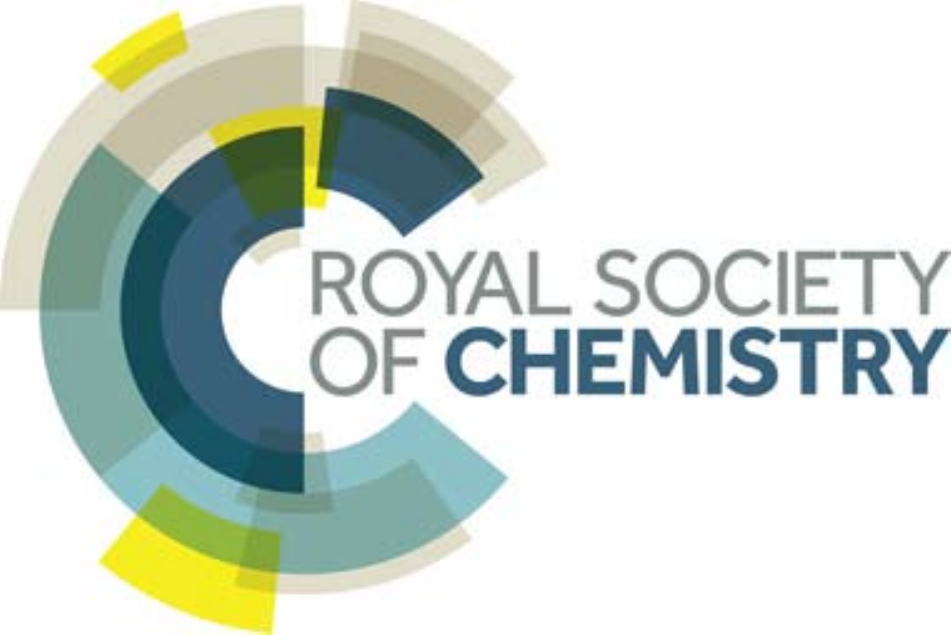}}
\renewcommand{\headrulewidth}{0pt}
}

\makeFNbottom
\makeatletter
\renewcommand\LARGE{\@setfontsize\LARGE{15pt}{17}}
\renewcommand\Large{\@setfontsize\Large{12pt}{14}}
\renewcommand\large{\@setfontsize\large{10pt}{12}}
\renewcommand\footnotesize{\@setfontsize\footnotesize{7pt}{10}}
\makeatother

\renewcommand{\thefootnote}{\fnsymbol{footnote}}
\renewcommand\footnoterule{\vspace*{1pt}%
\color{cream}\hrule width 3.5in height 0.4pt \color{black}\vspace*{5pt}} 
\setcounter{secnumdepth}{5}

\newcommand{\mbs}[1]{\bm{#1}}
\newcommand{\reals}{{\mathbb R}}

\makeatletter 
\renewcommand\@biblabel[1]{#1}            
\renewcommand\@makefntext[1]%
{\noindent\makebox[0pt][r]{\@thefnmark\,}#1}
\makeatother 
\renewcommand{\figurename}{\small{Fig.}~}
\sectionfont{\sffamily\Large}
\subsectionfont{\normalsize}
\subsubsectionfont{\bf}
\setstretch{1.125} 
\setlength{\skip\footins}{0.8cm}
\setlength{\footnotesep}{0.25cm}
\setlength{\jot}{10pt}
\titlespacing*{\section}{0pt}{4pt}{4pt}
\titlespacing*{\subsection}{0pt}{15pt}{1pt}

\fancyfoot{}
\fancyfoot[LO,RE]{\vspace{-7.1pt}\includegraphics[height=9pt]{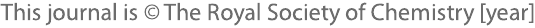}}
\fancyfoot[CO]{\vspace{-7.1pt}\hspace{13.2cm}\includegraphics{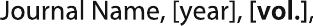}}
\fancyfoot[CE]{\vspace{-7.2pt}\hspace{-14.2cm}\includegraphics{head_foot/RF}}
\fancyfoot[RO]{\footnotesize{\sffamily{1--\pageref{LastPage} ~\textbar  \hspace{2pt}\thepage}}}
\fancyfoot[LE]{\footnotesize{\sffamily{\thepage~\textbar\hspace{3.45cm} 1--\pageref{LastPage}}}}
\fancyhead{}
\renewcommand{\headrulewidth}{0pt} 
\renewcommand{\footrulewidth}{0pt}
\setlength{\arrayrulewidth}{1pt}
\setlength{\columnsep}{6.5mm}
\setlength\bibsep{1pt}

\makeatletter 
\newlength{\figrulesep} 
\setlength{\figrulesep}{0.5\textfloatsep} 

\newcommand{\topfigrule}{\vspace*{-1pt}%
\noindent{\color{cream}\rule[-\figrulesep]{\columnwidth}{1.5pt}} }

\newcommand{\botfigrule}{\vspace*{-2pt}%
\noindent{\color{cream}\rule[\figrulesep]{\columnwidth}{1.5pt}} }

\newcommand{\dblfigrule}{\vspace*{-1pt}%
\noindent{\color{cream}\rule[-\figrulesep]{\textwidth}{1.5pt}} }

\makeatother

\twocolumn[
  \begin{@twocolumnfalse}
\vspace{3cm}
\sffamily
\begin{tabular}{m{4.5cm} p{13.5cm} }

\includegraphics{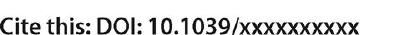} & \noindent\LARGE{\textbf{Influence of surface tension in the surfactant-driven fracture of closely-packed particulate monolayers$^\dag$}} \\
\vspace{0.3cm} & \vspace{0.3cm} \\

 & \noindent\large{Christian Peco \textit{$^{a}$},  Wei Chen \textit{$^{b}$}, Yingjie Liu \textit{$^{a}$}, M. M. Bandi \textit{$^{c}$}, John E.\ Dolbow \textit{$^{a}$}, and Eliot Fried \textit{$^{b}$}} \\

\includegraphics{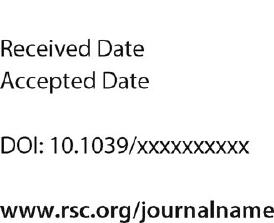} & \noindent\normalsize{	

		
	 A phase-field model is used to capture the surfactant-driven formation of fracture patterns in particulate monolayers.  The model is intended for the regime of closely-packed systems in which the mechanical response of the monolayer can be approximated as a linearly elastic solid.  The model approximates the loss in tensile strength of the monolayer as the surfactant concentration increases through the evolution of a damage field.  Initial-boundary value problems are constructed and spatially discretized with finite element approximations to the displacement and surfactant damage fields.   A comparison between model-based simulations and existing experimental observations  indicates a qualitative match in both the fracture patterns and temporal scaling of the fracture process. The importance of surface tension differences is quantified by means of a dimensionless parameter, revealing thresholds that separate different regimes of fracture. These findings are supported by newly performed experiments that validate the model and demonstrate the strong sensitivity of the fracture pattern to  differences in surface tension.  
	 
	 
	 } \\
\end{tabular}

 \end{@twocolumnfalse} \vspace{0.6cm}

  ]

\renewcommand*\rmdefault{bch}\normalfont\upshape
\rmfamily
\section*{}
\vspace{-1cm}


\footnotetext{\textit{$^{a}$~Department of Civil and Environmental Engineering, Duke University, Durham, NC 27708, USA}}
\footnotetext{\textit{$^{b}$~Mathematical Soft Matter Unit, OIST Graduate University, Onna-son, Okinawa, 904-0495, Japan }}
\footnotetext{\textit{$^{c}$~Collective Interactions Unit, OIST Graduate University, Onna-son, Okinawa, 904-0495, Japan }}





\section{Introduction}

When a densely packed monolayer of hydrophobic particles is placed on the surface of a liquid, the particles interact through capillary bridges that form,\cite{Varshney2012} leading to the formation of particle rafts.  The macroscopic properties of these rafts reflect an interplay between fluid and solid mechanics,\cite{Pickering1907, Binks2001, Cicuta2003} giving rise to novel physics. This interplay is relevant to a wide range of applications, from the synthesis of ``liquid marbles''~\cite{aussillous2001liquid} to the design of drug delivery systems~\cite{Tsapis2002} to the stabilization of drops.\cite{subramaniam2005}





The interest in particle rafts has driven researchers to investigate their mechanical properties.\cite{Cicuta2003,Vella2004} It is now known that densely packed monolayers exhibit a two-dimensional linearly elastic solid response, and that the mechanical properties of such monolayers depend proportionally on the surface tension of the liquid layer.    The introduction of a controlled amount of surfactant generates a surface tension gradient, producing Marangoni forces~\cite{Scriven1960} and causing the surfactant to spread, fracturing the monolayer. Studies of surfactant-driven fracture have examined the role of viscosity and the initial packing fraction on the evolution of cracks in closely and loosely-packed systems, respectively.\cite{Vella2006, bandi2011} Surprisingly, the potentially important role of differences in the surface tension of the surfactant and the underlying liquid has not been explored. The magnitude of this difference is interesting because it determines the Marangoni force exerted on the particulate monolayer and it is the main driving force for the fracture process.  Modulating the surface tension difference also provides a way to probe mechanical properties that are difficult to measure directly, such as the critical failure stress. Finally, the surface tension difference can easily be controlled in the laboratory by modifying the composition of the surfactant or the underlying liquid.

In this article, we focus on closely-packed systems. We develop a phase-field model that takes into account a two-way coupling between the flow of surfactant and the motion of the monolayer.  Through model-based simulations and accompanying experiments, we demonstrate that surface tension differences play a vital role in the overall fracture response of particle rafts.

\begin{figure}[h!]
	\centering
	\includegraphics[trim=0cm 0cm 0cm 0cm,height=2.6cm]{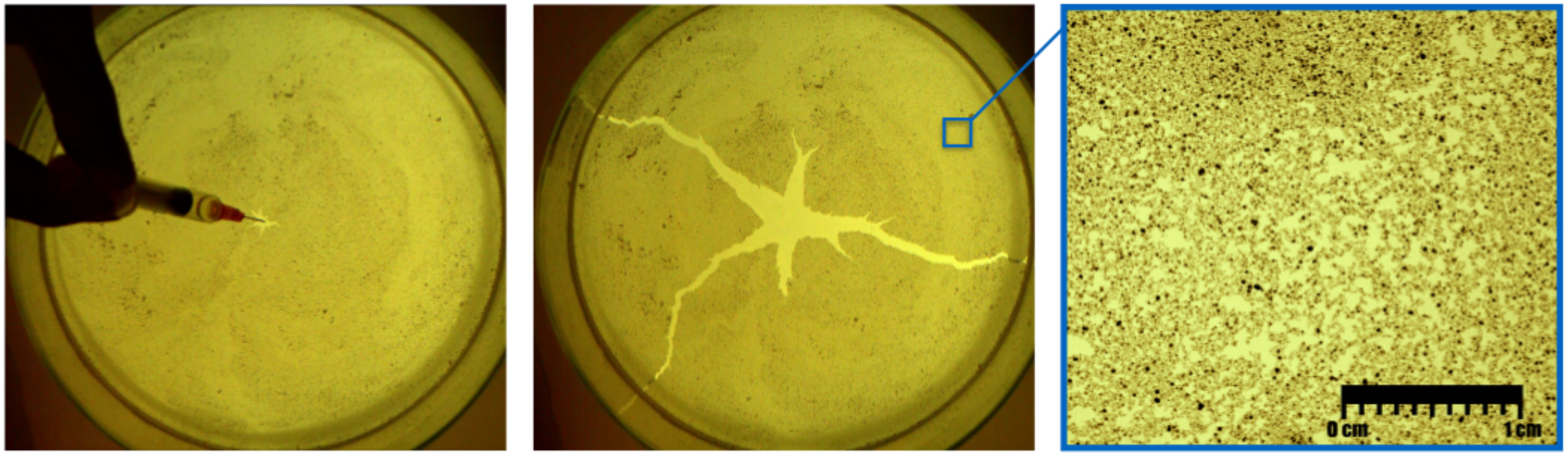}
	\caption{Left, center: States before and after the introduction of surfactant. The surfactant is injected at the center, decreasing the surface tension and generating an advancing front at which fracture initiates. Right: micrograph showing the microstructure of the closely-packed particulate monolayer.}
	\label{fig:intro}
\end{figure}

The general setup for a surfactant injection experiment, illustrated in Figure~\ref{fig:intro}, consists of a circular Petri dish containing a liquid layer onto which hydrophobic particles are deposited. The surfactant is introduced with a needle near the center of the dish. The surface tension of the liquid-vapor interface decreases where the surfactant is present. Marangoni forces then cause the surfactant to spread over the surface of the liquid and through the monolayer. The subsequent response is sensitive to the ratio of the fraction of the area of the liquid-vapor interface that is occupied by particles, which we refer to as the packing fraction and denote by $\phi$. As $\phi$ is increased, the properties of the liquid-vapor interface change from liquid to solid.\cite{bandi2011}

Consistent with out interest in closely-packed systems, we consider situations in which the packing fraction is high ($0.7 \leq \phi \leq 0.9$). We model the particle laden liquid-vapor interface as a continuous two-dimensional linearly elastic solid, capable of supporting both tension and compression.\cite{Cicuta2003, Vella2004} The mechanical properties of such a particulate monolayer~\cite{cicuta2009} are basic to understanding its response to surfactant-driven stresses. An estimate of those properties is provided by Vella et al.,\cite{Vella2004} based on experimental measurements and geometrical arguments. Surface wave experiments have been used to characterize the stretching and bending stiffnesses of particulate monolayers,\cite{Planchette2012} which are found to present a soft granular character with nonclassical response under fluid-driven deformation.\cite{MacMinn2015}

Experiments show that when such a system is stimulated by the localized introduction of a surfactant, an advancing front forms and fractures the monolayer.\cite{Vella2006} 
This type of surfactant-induced effect has also been observed in other systems, such as agar gels.\cite{Spandagos2012} The resulting fracture patterns, illustrated in Figure~\ref{fig:intro}, are reminiscent of those observed in classical brittle materials but with significant differences. Bandi et al.~\cite{bandi2011} suggested that the fracture patterns can be very sensitive to variations in the initial distribution of particles. Additionally, in conventional elastic solids, crack branching is mostly associated with inertial effects and dynamic instabilities (e.g.,~bifurcations which occur as the crack tip velocity approaches 60\% of the Rayleigh wave speed~\cite{Yoffe1951}). In contrast, crack branching has been observed in particulate monolayers at crack speeds as low as 0.2\% of the shear wave speed.\cite{Vella2006}  Regarding the time scales of the fracture process, the most obvious distinction is that crack tip velocities do not appear to be influenced by the elastic properties of the monolayers. Rather, experimental observations suggest that the fracture time scale is mostly governed by variations in the viscosity of the underlying liquid~\cite{Vella2006} and the packing fraction.\cite{bandi2011} These observations raise a number of questions regarding the fluid-driven fracture of elastic media in general and in monolayers in particular.

The mechanisms underlying the surfactant-driven fracture of particulate monolayers are challenging to study experimentally.  There are practical limits to the range of particle sizes that can be used.   Sufficiently small particles fall below the current resolution limit of imaging equipment when a full field of view is needed. Particles that are too large have too much inertia to move significantly in response to surfactant flow. 
Modeling and simulation efforts can address these concerns and provide detailed insight concerning the basic mechanisms and sensitivities. However, computational studies of these systems are in the early stage of development. Previous attempts to model particulate monolayers have focused on loosely-packed systems, which admit several simplifying assumptions.  For example, Bandi et al.,\cite{bandi2011} developed a discrete-element method to examine the influence of the initial packing fraction on the number of fractures in loosely-packed systems. In that work, a one-way coupling in which surfactant flow influences the motion of the particles, but not vice-versa, as assumed.  
While simulations based on this approach accurately reproduce the limiting number of cracks that develop, the underlying assumptions limit its applicability to situations where the packing fraction is low. 

The numerical simulation of closely-packed systems is challenging due to the high number of particles and the complexity of the physics involved in fracture. In this work, we propose a model based on a phase-field method which makes it possible to smoothly represent transitions between the damaged and undamaged zones of the monolayer as the surfactant advances. Phase-field models are suitable for this kind of problem since they avoid the need to track the propagating front, while allowing for a simple and powerful way to introduce the essential physics. We take as a starting point the work of Miehe et al.~\cite{Miehe2010a} and 
Borden et al.~\cite{Borden2012} for phase-field regularizations of the Griffith model for fracture.   Based on these approaches, we present a new model that includes several features that are important to characterize the fracture of particulate monolayers. In particular, our model incorporates the distribution of particles, the force on the monolayer due to the presence of surfactant, and the viscosity of the underlying liquid.  
Our objective is to develop the simplest model capable of capturing the salient features of this system, namely the sensitivity of fracture patterns to differences in surface tension and the temporal scaling of the fracture process.   

This paper is organized as follows. In Section~\ref{sect:cont}, we present a phase-field  approach for modeling the fracture mechanics of particulate monolayers, placing emphasis on the fundamental nature of the terms used to describe each contribution to the physics, after which we propose a governing free energy per unit area from which the model is derived. In Section~\ref{sect:exp}, we describe the materials and experimental methodology used to explore the fidelity of the computational model. In Section~\ref{sect:num}, we present numerical results which demonstrate the capability of the model to reproduce different cases of surfactant-driven fracture. We then present a phase diagram which delineates different fracture regimes as a function of the surface tension difference and fracture resistance of the monolayer.   Finally, in Section~\ref{sect:conc},  we summarize our main results and propose directions for further research.


\section{Phase-field model}
\label{sect:cont}

We model the surfactant-driven fracture of particulate monolayers by adapting recently developed phase-field approaches to fracture.\cite{Miehe2010a,Borden2012,Silva2013}  The model is designed to capture how a drop of surfactant introduced at the center of the domain can spread to form cracks or fissures in the particulate monolayer (Figure~\ref{fig:intro}). 
Accordingly, our model incorporates a number of postulates that are based on experimental observations, as detailed below.   Central among these  is the assumption that the fractured zones are completely filled with surfactant, without appreciable penetration of the surfactant beyond the contours of the cracks.  On this basis, we use a single ``surfactant damage'' field as an indicator function for the surfactant concentration and for the damage to the monolayer.  

\begin{figure}[b!]
	\centering
	\begin{overpic}[width=0.5\textwidth]{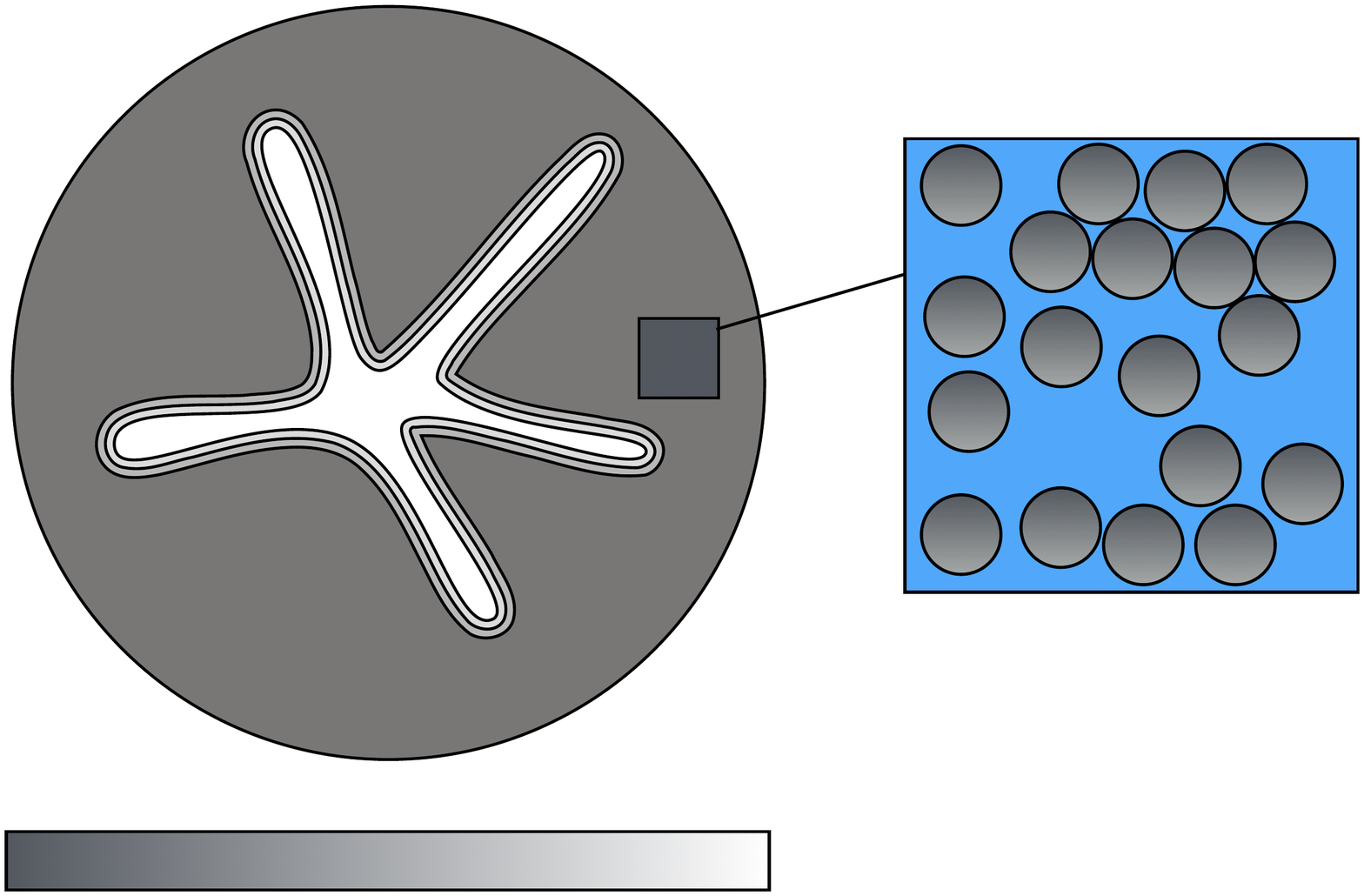}
		\put (28,2) {$d$}
		\put (5,7.5) {\color{white}$0.0$}
		\put (50,7.5) {$1.0$}
		\put (35,62) {\color{white}$\Omega$}
		\put (1,62) {a)}
		\put (60,62) {b)}
	\end{overpic}\vspace{-0.25cm}
	\caption{ (a) Schematic of a fractured particulate monolayer. The fractured zone is represented as a smooth transition zone across which the damage field $d$ takes values between zero and unity. (b) The microstructure of the elastic domain $\Omega$ is captured by an additional packing fraction field $\phi$, representing the solid to fluid ratio of the medium.}
	\label{fig:phasefield}
\end{figure}

\subsection{General considerations}
\label{sect:cont0}


Consider a monolayer of particles floating on the surface of a liquid layer. Let the two-dimensional region occupied by that monolayer be denoted by $\Omega$. In  Figure~\ref{fig:phasefield}(a), the dark portion of $\Omega$ represents the intact portion of the monolayer and the light portion of $\Omega$ represents the portion of the monolayer damaged by the surfactant. The state of the monolayer is described by two independent variables, its vector displacement field $\mbs{u}$ and a scalar surfactant damage field $d$. As indicated in Figure~\ref{fig:phasefield}(a), $d$ takes values in  $[0,1]$, with $d=1$ representing completely damaged material.
%
%
%
For closely-packed monolayers, experimental observations suggest that fracture occurs even for small values of the (infinitesimal) strain tensor
\begin{equation}
\mbs{\varepsilon}=\frac{1}{2}(\nabla\mbs{u}+(\nabla\mbs{u})^\trans\mskip1.5mu).
\end{equation}

The microstructure of the monolayer is described by a scalar packing fraction $\phi$ that takes values in $[0,1]$ and depends on position $\mbs{x}$ in $\Omega$, as illustrated in Figure~\ref{fig:phasefield}(b). This field characterizes the ratio of the subset of the surface of the liquid layer that is covered by particles to the total area of that surface and is thought to influence the overall fracture pattern, crack kinking, and branching.\cite{bandi2011} We assume that the particles are rigid, so that any local contraction or expansion of the  monolayer is accommodated solely by variations of the surface area not covered by particles. Thus, given an initial packing fraction $\phi_0$, the packing fraction $\phi$ depends on the trace, $\tr\mbs{\varepsilon}$, of the strain $\mbs{\varepsilon}$ through
\begin{equation}\label{eq:mob_update}   
\phi(\mbs{\varepsilon})=\frac{\phi_0}{1+\tr\mbs{\varepsilon}}.
\end{equation}
 

We further assume that surfactant damage acts only to degrade the tensile resilience of the monolayer and that crack propagation is prohibited under compression. This is achieved by employing a spectral decomposition~\cite{Miehe1998, Miehe2010a} of $\mbs{\varepsilon}$ into positive and negative components $\mbs{\varepsilon}_+$ and $\mbs{\varepsilon}_-$. It is then possible to define tensile and compressive strain-energy densities $W_+$ and $W_-$ through
\begin{equation}
W_\pm(\mbs{\varepsilon})=\frac{E}{2(1+\nu)}|\mbs{\varepsilon}_\pm|^2+\frac{\nu E}{(1-\nu)^2}(\tr_{\pm} \mbs{\varepsilon})^2,
\end{equation}
where $E>0$ and $0<\nu<1$ are the Young's modulus and Poisson's ratio of the undamaged monolayer and the positive and negative trace operations $\tr_+$ and $\tr_-$ are defined in accord with
\begin{equation}
\tr_+\mbs{\varepsilon} 
=\begin{cases}
\tr \mbs{\varepsilon}, & \mbox{if} \quad\tr \mbs{\varepsilon} \ge 0, 
\cr\noalign{\vskip4pt}
0, & \mbox{otherwise},
\end{cases}
\end{equation}
and
\begin{equation}
\tr_-\mbs{\varepsilon}
=\begin{cases}\tr\mbs{\varepsilon}, & \mbox{if} \quad\tr \mbs{\varepsilon} \le 0,
\cr\noalign{\vskip4pt}
0, & \mbox{otherwise}.
\end{cases}
\end{equation}



%
%

We assume that the free-energy density of the monolayer is a function $\psi$ of $\mbs{\varepsilon}$, $d$, and $\nabla d$ with the  form
\begin{equation}
\label{eq:energy}
\psi(\mbs{\varepsilon},d,\nabla d)=\underbrace{(1-d)^2W_+(\mbs{\varepsilon})}_\text{tensile}
+\underbrace{\xi(\phi(\mbs{\varepsilon})) W_-({\mbs{\varepsilon}})}_\text{compressive}
+\underbrace{Q(d, \nabla d)}_\text{fracture}
+\underbrace{F ({\mbs{\varepsilon}}, d)}_\text{surfactant}.
\end{equation}
As in conventional phase-field models for brittle fracture,\cite{Miehe2010a,Borden2012}  the tensile contribution to $\psi$ decays quadratically with the surfactant damage $d$. The remaining contributions of the energy are, however, nonstandard and therefore require further discussion.





\begin{figure}[t!]
	\centering
	\begin{overpic}[trim=0cm -3cm 0cm 0cm,width=0.25\textwidth]{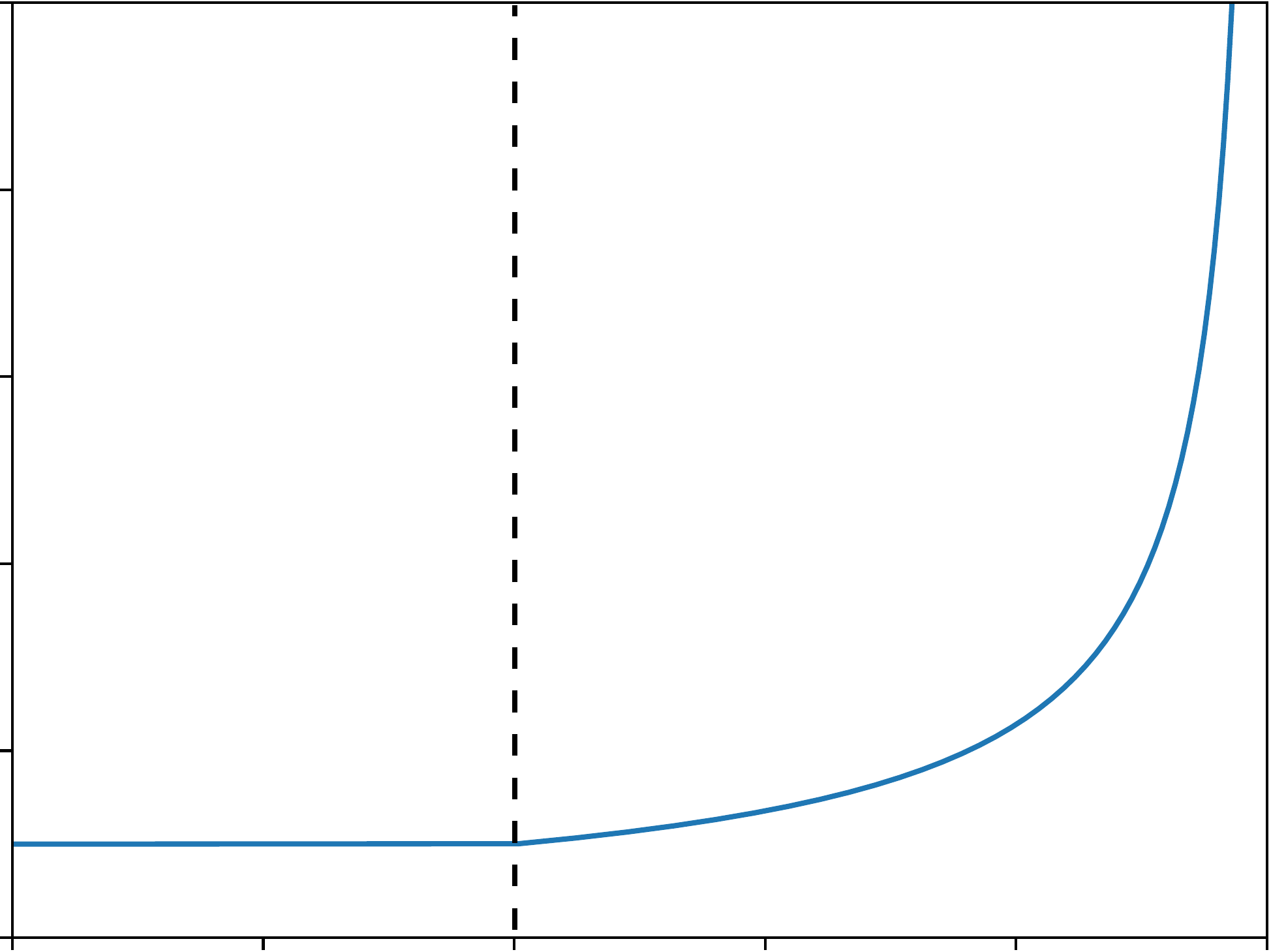}


\put (-5,14) {$0$}
\put (-5,29) {$2$}
\put (-5,44) {$4$}
\put (-5,58) {$6$}
\put (-5,73) {$8$}
\put (-8,87) {$10$}

\put (-15,51) {$\xi$}
\put (15,25) {$1.0$}

\put (-5,8) {$0.80$}
\put (15,8) {$0.82$}
\put (34,8) {$0.84$}
\put (53,8) {$0.86$}
\put (72,8) {$0.88$}
\put (91,8) {$0.90$}

\put (47,0) {$\phi$}

\end{overpic}
	\caption{Variation in the jamming factor $\xi$ with packing fraction $\phi$. The jamming factor tends towards $\infty$ as $\phi$ surpasses the jamming threshold $\phi_j=0.84$ and approaches the maximum packing fraction $\phi_m=0.9$.}
	\label{fig:features2}
\end{figure}
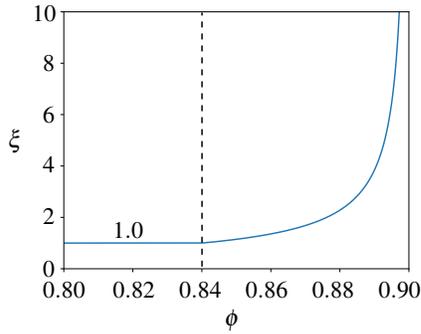

The compressive contribution to $\psi$ accounts for increases in compressive energy that accompany increases in the packing fraction through the jamming factor $\xi$. This factor penalizes compression beyond a jamming threshold $\phi_j$ and prevents packing fractions from exceeding a maximum value $\phi_m$. The variation of $\xi$ with $\phi$ --- and, thus, with reference to \eqref{eq:mob_update}, the initial packing fraction $\phi_0$ and strain $\mbs{\varepsilon}$ --- is illustrated in Figure~\ref{fig:features2}. Its value starts at $\xi=1$ for $0\le\phi\le\phi_j$ and increases monotonically for $\phi_j\le\phi<\phi_m$, exhibiting a vertical asymptote as $\phi\to\phi_m$. The thresholds $\phi_j=0.84$ and $\phi_m=0.9$ correspond to a random close-packing in two space dimensions\cite{OHern2003} and the maximum packing density for two-dimensional discs, respectively. Numerically, it is more convenient to use expressions that effectively raise the compressive contribution without becoming unbounded. In this work, we use an expression with the form $1.0 + f(1.0+\text{tanh}((\phi-0.5(\phi_m-\phi_j))/l)$, with $f$ being the factor of amplification and $l$ being the length scale controlling the width of the regularization.



For the contribution to $\psi$ associated with fracture, we choose a modified version 
\begin{equation}\label{eq:frac_nrg1}
Q(d, \nabla d)=\frac{G(d)}{2}\left(\frac{d^2}{\lambda}+\lambda|\nabla d|^2\right)
\end{equation}
of the expression proposed by Miehe et al.~\cite{Miehe2010a}, with $G$ defined by
\begin{equation}\label{eq:frac_nrg2}
G(d)=G_0\Big(\frac{\gamma_s}{\gamma_f}+\Big(1-\frac{\gamma_s}{\gamma_f}\Big)(1-d)\Big)
=G_0\Big(1-\Big(1-\frac{\gamma_s}{\gamma_f}\Big)d\Big),
\end{equation}
where $\gamma_f$ and $\gamma_s$ are the surface tension of the liquid layer and the surfactant, respectively, $G_0>0$ is a constant that represents the fracture toughness of the undamaged monolayer, and $\lambda>0$ is a constant proportional to the characteristic thickness of a layer between damaged and undamaged material. This modification accounts for a reduction in the fracture toughness with increasing surfactant concentration. Accordingly, it is convenient to introduce
\begin{equation}\label{eq:gcr}
G_r = G(1) =  G_0 \frac{\gamma_s}{\gamma_f}
\end{equation}
as a measure of the reduced fracture toughness of the monolayer. 

\begin{figure}[t!]
		\centering
		\includegraphics[trim=2cm 0cm 3cm 1cm,height=2cm]{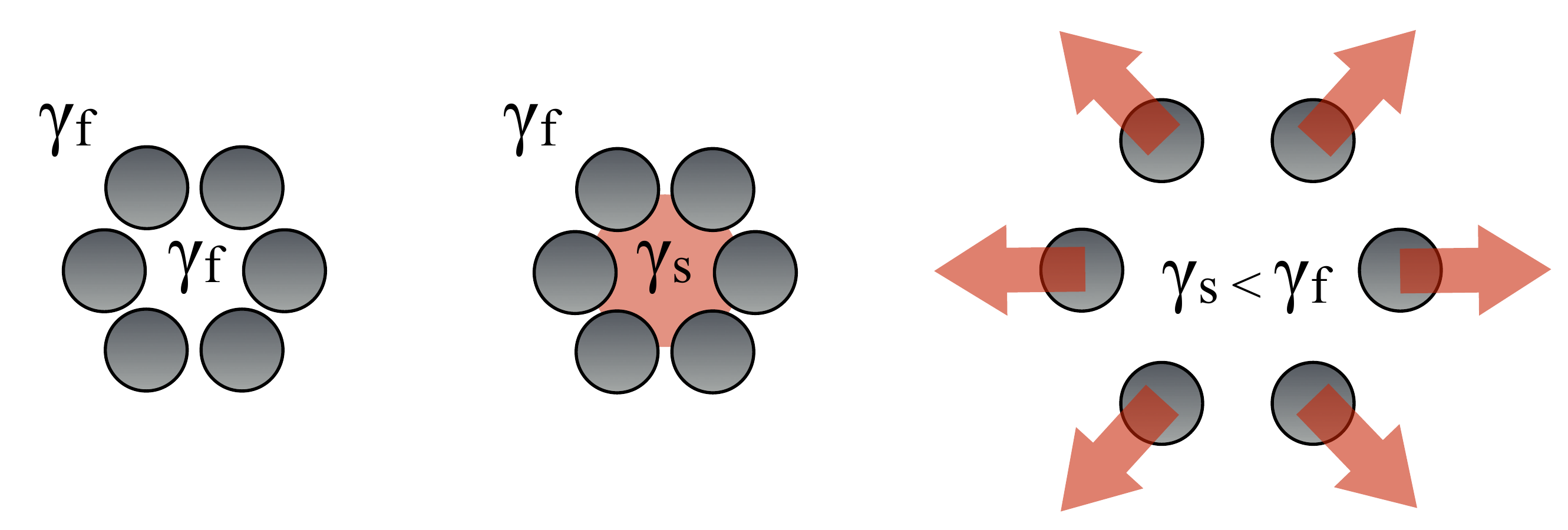}
		\caption{Representation of the surfactant force driving the system. A collection of particles, initially in equilibrium at a constant surface tension $\gamma_f$, are reached by the surfactant. The gradient between its lower surface tension $\gamma_s$ and the surrounding media generates a Marangoni flow, which pushes the particulate monolayer outward.}
		\label{fig:features1}
	\end{figure}

Finally, the surfactant contribution to $\psi$ accounts for the interplay between the monolayer and the surfactant (Figure~\ref{fig:features1}) and is assumed to be of the form
\begin{equation}
F(\mbs{\varepsilon},d)=\frac{F_0(\gamma_f-\gamma_s)(\tr\mbs{\varepsilon})d^2}{2},
\label{F}
\end{equation}
where $F_0>0$ is a constant.

%
%
%
%
%
%
%
%

\subsection{Governing equations}
	
Following Silva et al.\cite{Silva2013} (but neglecting inertia), the governing equations of the phase-field model consist of the macroforce balance
\begin{equation}
\text{div}\Big(\frac{\partial \psi({\mbs{\varepsilon}},d,\nabla d)}{\partial {\mbs{\varepsilon}}}\Big)={\bf0}
\label{fb}
\end{equation}
and the microforce balance
\begin{equation}\label{eq:cases}
\text{div}\Big(\frac{\partial \psi({\mbs{\varepsilon}},d,\nabla d)}{\partial \nabla d}\Big)
-\frac{\partial \psi({\mbs{\varepsilon}},d,\nabla d)}{\partial d}
=\begin{cases}
\mskip8.25mu\beta\dot{d}, & \text{if} \quad \dot{d}>0, 
\\ 
-\pi_r, & \text{if} \quad \dot{d} = 0,
\end{cases}
\end{equation}
where $\beta\ge0$ represents the kinetic modulus that controls the rate at which cracks can propagate through the monolayer. The alternative on the right-hand side of \eqref{eq:cases} embodies the requirement that, consistent with experimental observations, cracks that form in the monolayer never heal. This requirement takes the form of the constraint $\dot{d}\ge0$, and a  reactive microforce $\pi_r$ is needed to ensure satisfaction of that constraint. In particular, $\pi_r$ vanishes for $\dot{d} > 0$ and is determined by the left-hand side of \eqref{eq:cases} for $\dot{d}=0$.

The kinetic modulus $\beta$ accounts for two effects.   First, it captures the capacity of the surfactant to penetrate the particulate monolayer and thereby generate damage, which is influenced by $\phi$ and the radius of the particles. As $\phi$ and the particle size increase, the higher density of particles and the increased tortuosity impede surfactant spreading. Second, it captures the resistance of the underlying liquid layer to  rearrangements of the particles, a resistance which is directly related to the viscosity $\mu_f$ of the liquid comprising that layer. For the foregoing reasons, we assume that $\beta$ increases with $\phi$, $\mu_f$, and the mean particle radius $\bar{r}_p>0$ in accord with the relation 
\begin{equation}\label{eq:mobility} 
\beta=\beta_0\frac{1-\phi_m}{1-\phi},\qquad \beta_0=\frac{C\mu_f\bar{r}_p}{1-\phi_m},
\end{equation}
where $C>0$ is a constant and $\beta_0$ represents the kinetic modulus for the case of maximum packing $\phi=\phi_m$.



Since the kinetic modulus $\beta$ defined in \eqref{eq:mobility} is positive, damage only increases when the left-hand side of \eqref{eq:cases} is positive. Accordingly, the evolution equation \eqref{eq:cases} for $d$ becomes
\begin{equation}
\frac{1-\phi(\mbs{\varepsilon})}{\beta_0(1-\phi_m)}\Big\langle 
\text{div}\Big(\frac{\partial \psi({\mbs{\varepsilon}},d,\nabla d)}{\partial \nabla d}\Big)
-\frac{\partial \psi(\mbs{\varepsilon},d,\nabla d)}{\partial d}\Big\rangle=\dot{d},
\label{mfb}
\end{equation}
where, given a scalar-valued quantity $h$, 
\begin{equation}
\langle h\rangle=\begin{cases}
0, & h \le 0, 
\\ 
h, & h > 0, 
\end{cases}
\end{equation}
denotes its Macaulay bracket.

With reference to the right-hand side of the definition \eqref{eq:energy} of the free-energy density $\psi$ and using  \eqref{eq:frac_nrg1}--\eqref{F} and \eqref{eq:mobility}, we find the governing equations \eqref{fb} and \eqref{mfb} can be written as
\begin{equation}\label{eq:surf_evol0}
\text{div}\Big((1-d)^2\frac{\partial W_+(\mbs{\varepsilon})}{\partial\mbs{\varepsilon}}+\frac{\partial (\xi(\phi(\mbs{\varepsilon}))W_-(\mbs{\varepsilon}))}{\partial {\mbs{\varepsilon}}}+\frac{F_0(\gamma_f-\gamma_s)d^2}{2}\mbs{I}\Big)={\bf0}
\end{equation}
and
\begin{multline}
\frac{1-\phi(\mbs{\varepsilon})}{\beta_0(1-\phi_m)}\Big\langle 
G_0\lambda\Big(1-\Big(1-\frac{\gamma_s}{\gamma_f}\Big)d\Big)\Delta d-\frac{G_0\lambda}{2}\Big(1-\frac{\gamma_s}{\gamma_f}\Big)|\nabla d|^2
\\
+2(1-d)W_+(\mbs{\varepsilon})
-F_0(\gamma_f-\gamma_s)(\tr\mbs{\varepsilon})d
\\
-\frac{G_0d}{\lambda}\Big(1-\frac{3}{2}\Big(1-\frac{\gamma_s}{\gamma_f}\Big)d\Big)
\Big\rangle
=\dot{d},
\label{mfbdinal0}
\end{multline}
respectively, and where $\mbs{I}$ denotes the (two-dimensional) identity tensor. We consider \eqref{eq:surf_evol0} and \eqref{mfbdinal0} subject to the zero displacement condition
\begin{equation}
\mbs{u} = \mbs{0}
\end{equation}
on the boundary $\partial\Omega$ of $\Omega$ and, letting $\mbs{n}$ denote a unit normal on $\partial\Omega$, the natural boundary condition
\begin{equation}
\nabla d\cdot\mbs{n}=0.
\end{equation}
Additionally, we impose the initial conditions
\begin{equation}
\phi(x,0)=\phi_0(x), \qquad d(x,0)=d_0(x),
\end{equation}
with $\phi_0$ representing the initial distribution of particles and $d_0$ the initial damage.  

\begin{table}[h!]
\footnotesize
	\centering \caption{List of parameters in the model. }
	\label{table:modelparameters} 
	\begin{tabular}{p{2cm} p{5cm}} \hline
		parameter & description\\
		\hline
		$\phi_j$ & jamming packing fraction of monolayer\\
		$\phi_m$ & maximum packing fraction  of monolayer\\
		$\bar{r}_p$ & average particle radius\\
		$\gamma_{f}$ & surface tension of liquid layer\\
		$\gamma_s$ &  surface tension of surfactant\\
		$G_0$ & fracture toughness of undamaged monolayer    \\
		$F_0$ & surfactant force constant  \\
		$E$ & Young's modulus of monolayer\\
		$\nu$ & Poisson's ratio of monolayer\\
		$\mu_f$ & dynamic viscosity of monolayer\\
		$\beta_0$ & kinetic parameter\\
		\hline \end{tabular} \end{table}


\subsection{Model characterization: dimensionless number $\chi$}
\label{sect:cont2}




Introducing characteristic measures $L$ and $T$ of length and time, we define dimensionless counterparts $\mbs{x}^*$ and $t^*$ of $\mbs{x}$ and $t$ by
 \begin{equation}
\mbs{x}^*=\frac{\mbs{x}}{L}
\qquad\text{and}\qquad
t^*=\frac{t}{T}.
\label{starredindvbls}
\end{equation}
Thus, defining dimensionless versions
\begin{equation}
\nabla^* d = L \nabla d, 
\qquad 
\Delta^* d = L^2 \Delta d,
\qquad\text{and}\qquad 
\frac{\partial d}{\partial t^*} = T\dot{d}
\end{equation}
of the gradient, Laplacian, and partial time-derivative, dimensionless parameters,
\begin{equation}\label{eq:dim_params}
\left.
\begin{array}{c}
\displaystyle
G^*=\frac{G_r}{E \bar{r}_p},\qquad F^*=\frac{F_0(\gamma_f-\gamma_s)}{E}, \qquad L^*=\frac{L}{\bar{r}_p}, 
\cr\noalign{\vskip4pt}
\displaystyle
\beta_0^*= \frac{\beta_0}{ET}, \qquad \gamma^*=\frac{\gamma_s}{\gamma_f}, \qquad\text{and}\qquad \lambda^*=\frac{\lambda}{L},
\end{array}
\right\}
\end{equation}
and dimensionless strain-energy densities
\begin{equation}
W^*_\pm(\mbs{\varepsilon})=\frac{1}{2(1+\nu)}|\mbs{\varepsilon}_\pm|^2+\frac{\nu}{(1-\nu)^2} ( \tr_{\pm} \mbs{\varepsilon})^2,
\end{equation}
we arrive at dimensionless counterparts
\begin{equation}\label{eq:surf_evol0_nodim}
\text{div$^*$}\Big((1-d)^2\frac{\partial W^*_+(\mbs{\varepsilon})}{\partial\mbs{\varepsilon}}+\frac{\partial (\xi(\phi(\mbs{\varepsilon}))W^*_-(\mbs{\varepsilon}))}{\partial {\mbs{\varepsilon}}}+\frac{F^*}{2}d^2\mbs{I}\Big)={\bf0}
\end{equation}
and
\begin{multline}
\frac{1-\phi(\mbs{\mbs{\varepsilon}})}{\beta_0^*L^*(1-\phi_m)}
\Big\langle 
\frac{G^*}{\gamma*}\lambda^*(1-(1-\gamma^*)d)\Delta^* d-\frac{G^*}{\gamma*}\frac{\lambda^*}{2}(1-\gamma^*)|\nabla^* d|^2
\\
-\frac{G^*}{\gamma*}\frac{d}{\lambda^*}(1-\frac{3}{2}(1-\gamma^*)d)
\\
+2(1-d)L^*W^*_+(\mbs{\varepsilon})
-L^*F^*(\tr\mbs{\varepsilon})d
\Big\rangle
=\frac{\partial d}{\partial t^*}
\label{mfbdinal}
\end{multline}
of the governing equations \eqref{eq:surf_evol0} and \eqref{mfbdinal0}.

While we have not undertaken an exhaustive suite of simulations spanning the complete range of parameter space associated with the six dimensionless variables in \eqref{eq:dim_params}, the extensive testing that we have conducted indicates that the fracture pattern is largely governed by the dimensionless driving force $F^*$ and the dimensionless fracture toughness $G^*$  of the monolayer.   In particular, the number of fractures in the final configuration appears to be dictated by the ratio
\begin{equation}\label{eq:chi}
\chi = \frac{(F^*)^2}{G^*}.
\end{equation}
In Section~\ref{sect:relation}, we discuss the threshold levels of the dimensionless parameter $\chi$ that delineate different regimes, ranging from no fractures at all to a configuration with multiple branches.  



 The expression \eqref{eq:chi} for  $\chi$ can be further simplified by 
invoking the approximate scaling of  Young's modulus with the surface tension as derived by Vella et al.\cite{Vella2004} (i.e., $E \propto  \gamma_f/ \bar{r}_p$).
  The expression for $\chi$ can therefore be rewritten to isolate the influence of the surface tensions, yielding
\begin{equation}\label{eq:chi2}
\chi  \propto \frac{(F_0\bar{r}_p)^2}{G_0}\frac{(\gamma_f-\gamma_s)^2}{\gamma_s}.
\end{equation}
Daniels et al.\cite{Daniels2007} proposed a similar relationship between the elastic and Marangoni energies to explain fracture patterns in agar gels. In that work, the number of crack branches was found to scale linearly with the difference in surface tension.  As described in Section~\ref{sect:num}, our experimental observations and model-based simulations suggest the system considered here exhibits a much stronger sensitivity to surface tension differences.

\section{Experiments}
\label{sect:exp} 

\subsection{Materials}

All videos from experimental work were acquired using a Phantom V641 high-speed camera equipped with an AF Nikkor 50 mm f/1.8 D lens. Videos were saved using PCC software provided by Phantom. A 5 W white LED board used to illuminate the mixture was supplied by VANCO (series \#33342). Dispersion of silica microballoons (SIG MFG) was performed using the Active Motif Q120AM (120 W, 20 kHz) ultrasonicator equipped with a CL18 3.2 mm probe. Oleic acid (surfactant) was supplied by WAKO Chemicals and the acetone used as a cosolvent was supplied by Nacalai Tesque. The surfactant was dropped using Terumo 2.5 mL syringes (SS-02SZ) equipped with a 25 G (0.50 x 16 mm) needle. Surface tension measurements of mixtures were obtained using the KSV NIMA LB Small trough (33473).


\begin{figure}[tb!]
\centering
 \begin{overpic}[trim=4cm 3.7cm 3cm 1cm,height=6cm] {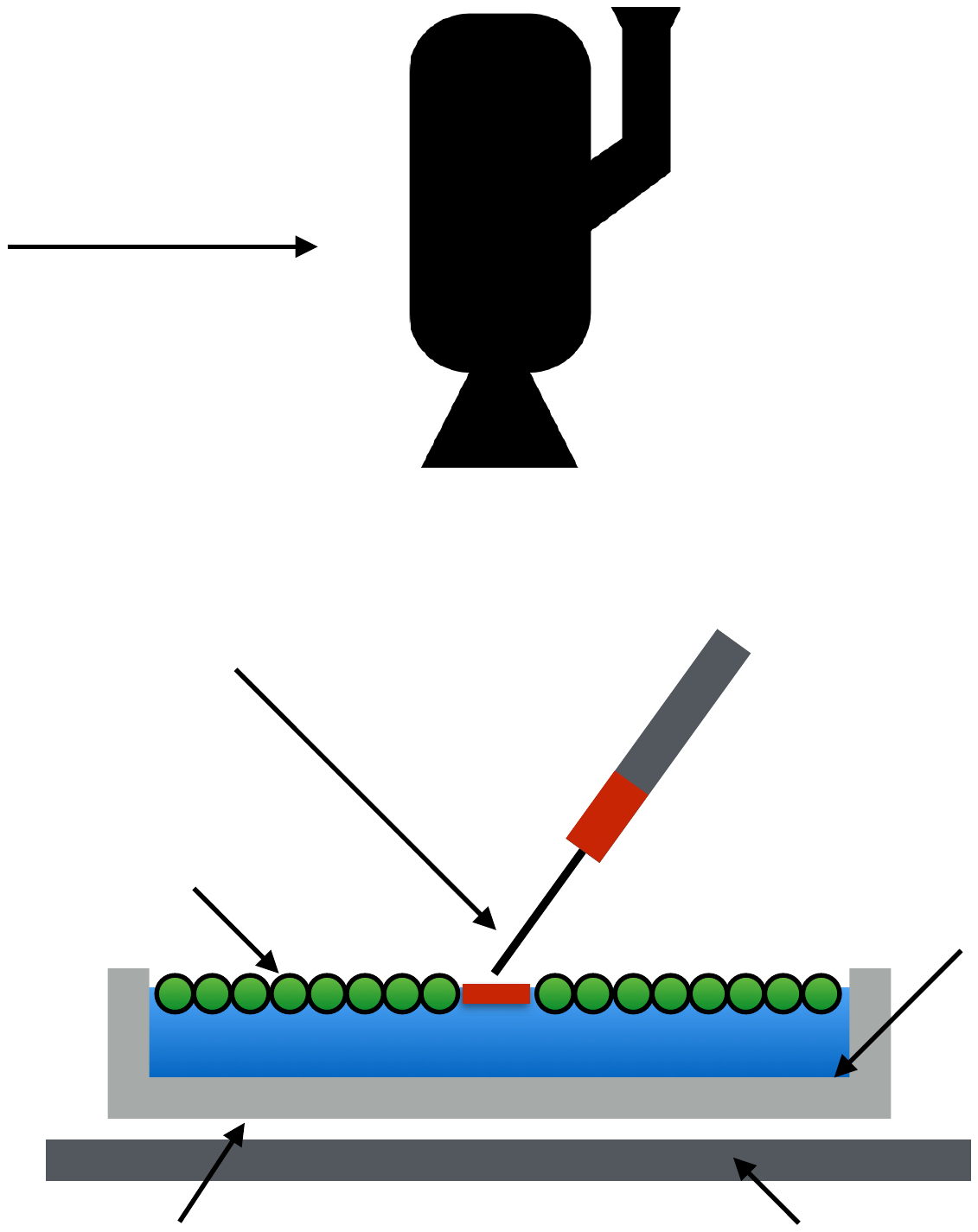}
 \put(11,53){\small  camera}
\put(12,33){\small  surfactant injection}
\put(25,21.5){\small  particles}
\put(64,18){\small  liquid layer}
\put(57,-0.5){\small  LED panel}
\put(23,-0.5){\small  Petri dish}
\end{overpic}
\caption{Setup: A clean low-form cylindrical vessel containing distilled water was placed on an LED tablet. A camera suspended vertically above records particulate monolayer compaction dynamics at the air-water interface when surfactant was introduced using a 25 G steel needle.}
\label{fig:exp}
\end{figure}

\subsection{Methodology}

Figure~\ref{fig:exp} shows a schematic of the experimental setup. A clean low-form cylindrical glass vessel 17 cm in diameter and 9 cm in height filled to a height of 2.5 cm with milli-Q water (or mixed with a cosolvent) was placed atop an LED light panel. Microballoons were carefully weighed to either 0.150 g or 0.250 g, introduced at the air-water interface, and dispersed by means of ultrasonication at 25\% power to form a particulate monolayer. Using a 2.5 mL syringe fitted with a 25 G metal tip, a single drop of surfactant (oleic acid with or without solvent) was introduced to the approximate center of the vessel at time t = 0 s. The water immiscible surfactant was observed to spread and push the microballoon particles radially outwards resulting in compaction of the particulate monolayers. Observations were captured using a Phantom high-speed camera at different frame rates. Data analysis was performed by first converting raw data files (.cine) to multipage image files (.tif) and these files were subsequently analyzed using ImageJ. In cases where the surfactant or the bulk water phase was mixed with a solvent to reduce surface tension, acetone was added carefully and aliquots of the liquids were quickly transferred to an LB trough for surface tension measurements.

\section{Results and discussion}
\label{sect:num}

\subsection{Pattern comparison}

We discretize the model described in Section~\ref{sect:exp} by recasting the evolution equations in variational form and approximating the displacement and surfactant damage fields with finite element basis functions.  Initial-boundary value problems are constructed over circular domains representative of the petri dishes used in the experiments.  
The experiments have a stochastic aspect corresponding to the initial particle locations.  This effect is modeled at the continuum scale by using initial packing fields with spatial variability.  Specifically, we numerically construct initial packing fraction fields over the circular domain that have a mean value of $\bar{\phi} = 0.795$ with a 6\% variation, corresponding to uniform random distributions with minimum and maximum packing fractions of 0.75 and 0.84, respectively.  
 The initial drop of surfactant is approximated by a sufficiently small (on the order of $\lambda$), empty circular region ($\phi_0=0$) concentric with the center of the dish.  
Table~\ref{table:parameters} provides the reference values of the parameters used for the simulations in this section. All simulations were performed using MOOSE (Multiphysics Object-Oriented Simulation Environment), a finite-element framework primarily developed by Idaho National Laboratory.\cite{Moose}

\begin{table}[tb!]
\footnotesize
\centering \caption{Dimensional variables.\cite{Vella2006, Vella2004}}
\label{table:parameters} 
\begin{tabular}{p{1.0cm} p{3.2cm}p{2.7cm}p{0.5cm}} \hline
 Property & Description & Value  & Unit \\
\hline
$r_d$  & radius of Petri dish  &  $8.5\cdot10^{-2}$   &      m	  \\
$ [ \phi^0(x) ]$ & range of initial packing fraction  &  $(7.5$--$8.4)\cdot10^{-1}$  &  none	\\
$\phi_m$ & packing fraction of jammed monolayer  &  $9.0\cdot10^{-1}$  & none   \\
$G_0$ & fracture toughness of undamaged monolayer &  $5.7\cdot(10^{-4}$--$10^{-2})$  & N/m     \\
$F_0$ & surfactant force constant &  $1.0\cdot(10^{-1}$--$10^{2})$  & 1/m\\
$\gamma_{f}$ & surface tension of liquid layer (water)  &  $7.2\cdot10^{-2}$  & N/m   \\
$\gamma_s$ & surface tension of surfactant (oleic acid) &  $4.0\cdot10^{-2}$  &  N/m \\
$\mu_{f}$ & range of dynamic viscosity of liquid layer &  $1.0\cdot(10^{-3}$--$10^{-1})$   & Pa s  \\
$\bar{r}_p$ & Mean particle radius  &  $1.0\cdot10^{-4}$   & m  \\
$E$ & Young's modulus of monolayer &  $3.178\cdot10^{3}$   & Pa \\
$\nu$ & Poisson's ratio of monolayer &  $1/\sqrt{3}$   & none  \\
$\beta_0$ & kinetic parameter  &  $1.0$   & 1/m  \\
$\lambda$  & characteristic length scale &  $1.0\cdot10^{-2}$   & 1/m  \\
\hline \end{tabular} \end{table}

\begin{table}[tb!]
\footnotesize
\centering \caption{Dimensionless parameters.}
\label{table:dimensionless} 
\begin{tabular}{p{1.0cm} p{2.0cm}} \hline
 Parameter & Value \\
\hline
$G^*$ & $10^{-3}$--$10^{-1}$  \\
$F^*$ & $10^{-4}$--$10^{-1}$  \\
$L^*$ & $8.5\cdot10^2$  \\
$\beta_0^*$ & $2.7\cdot(10^{-8}$--$10^{-6})$ \\
$\gamma^*$ & $5.6\cdot10^{-1}$ \\
$\lambda^*$ & $1.5\cdot10^{-1}$ \\
\hline \end{tabular} \end{table}



\begin{figure}[tb!!]
   \centering
   \begin{overpic}[trim=6.5cm 0cm 3cm 0cm,height=5cm]{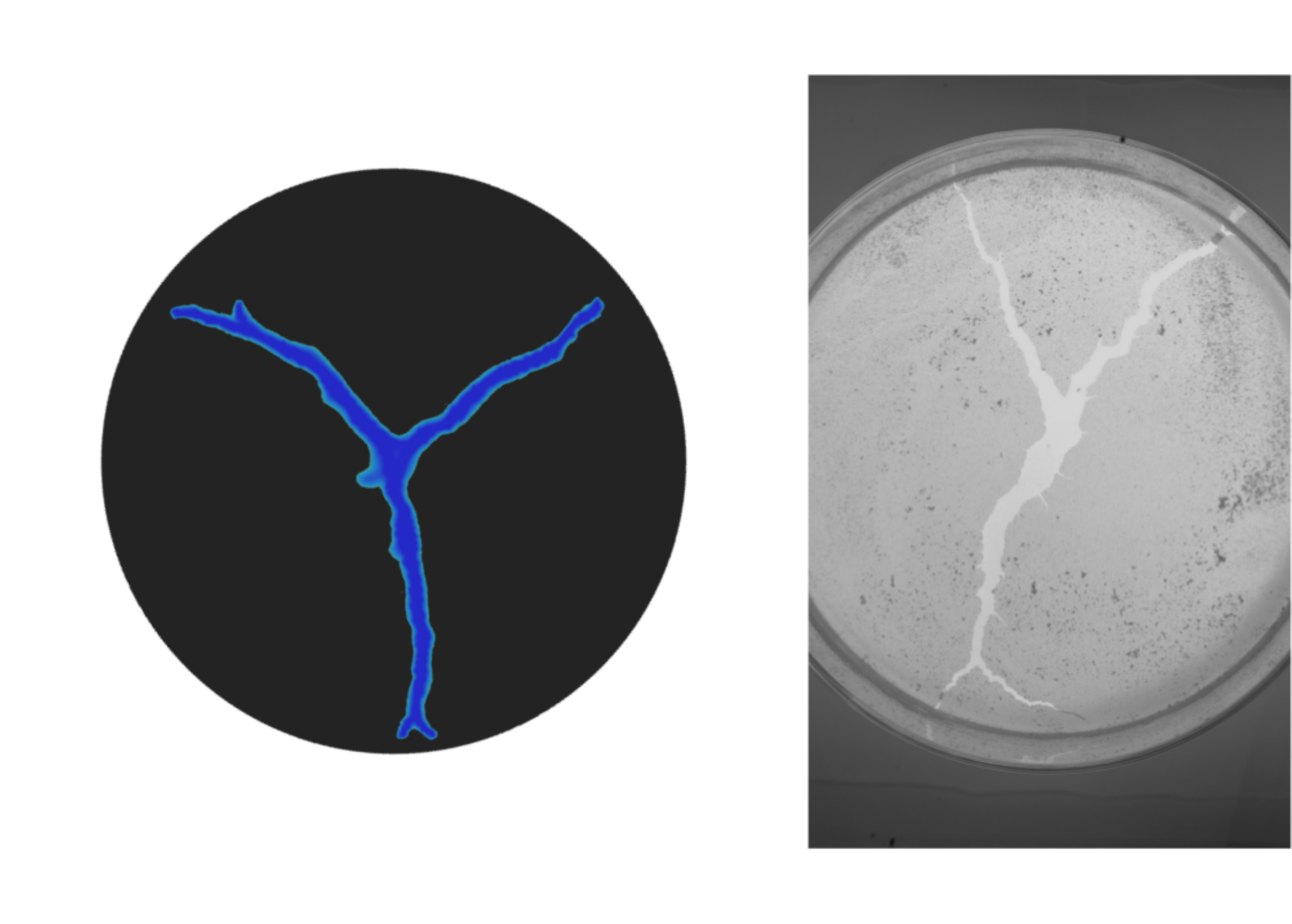}
       \end{overpic}\vspace{-0.25cm}
   \caption{Comparison between a simulation and a experimental observation for a three-branch fracture pattern. The model captures the main traits of the final configuration, as the number of fractures, and more detailed features corresponding to variations in the distribution of particles, such as secondary fracturing, crack bending, and branching. }
    \label{fig:comp1}

\end{figure}

A first qualitative assessment of the model is shown in Figure~\ref{fig:comp1}. In this example, the parameter $\chi$ is adjusted to reproduce a three-branch pattern, which is a commonly observed configuration in the experiments for our given set of parameters. The result indicates that the model captures the general fracture process and reproduce the final three-branch configuration. Additionally, the cracks are sensitive to the packing fraction structure, which gives rise to more subtle features. These features include the particular angle in which the three-branch pattern is generated, the kinking and bending of cracks when gradients in the particle density are encountered, and the bifurcation of a crack when it encounters a high density region directly in line with its propagation. In this work, stochastic features of the experiments are approximated through random variations in the initial packing fraction field. The specific details of the numerically generated fracture patterns thus stem from the initial conditions and are sensitive to the particular random seed used in the simulations.  As a result, our simulation results are only designed to capture the generic features of the experiments. Nevertheless, the model reproduces the number of cracks observed in the experiments and, at a qualitative level, the features of crack kinking and the formation of secondary branches.

\begin{figure}[tb!]
   \centering
   \includegraphics[trim=0cm 0cm 0cm 0cm,height=4.3cm]{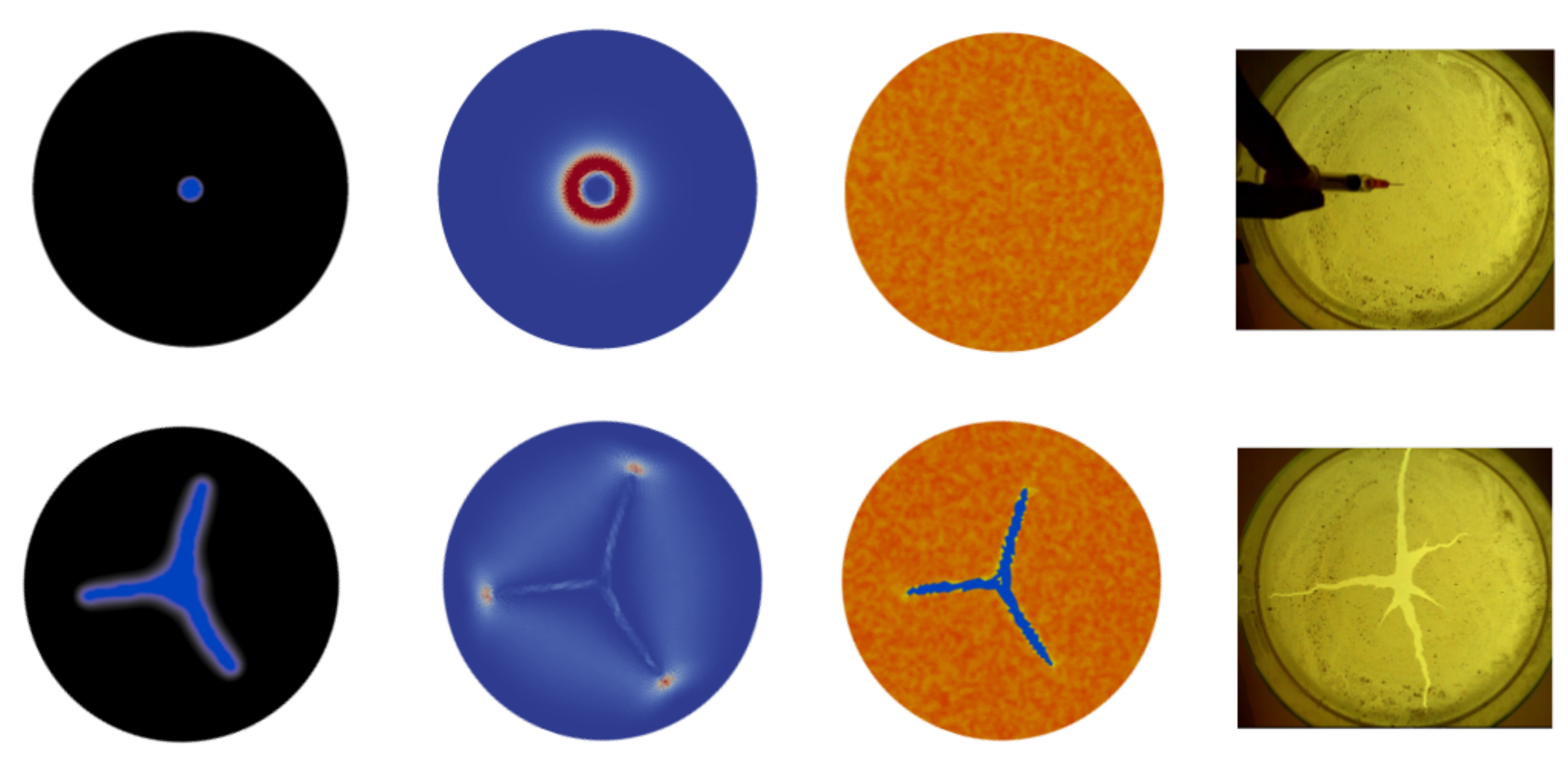}
   \caption{Numerical fields obtained for a simulation of a three-branch fracture pattern. From left to right, each column shows the initial (top) and final (bottom) states of surfactant damage $d$, tensile energy $W^+$ and $\phi$ fields. The rightmost column shows the reference experiment, obtained by adding oleic acid to a 0.25 g mass monolayer of particles laying on a 18\% salt saturation water solution.}
    \label{fig:comp3}
\end{figure}

We now provide a detailed description of a representative simulation in which a three-branch fracture pattern arises.  Figure~\ref{fig:comp3} shows the results from a simulation of this common pattern, obtained with the parameters given in Table~\ref{table:parameters} and $\chi=0.0125$  (see Table~\ref{table:dimensionless} for the correspondent dimensionless quantities). The columns in Figure~\ref{fig:comp3} show the evolution of three fields, namely the surfactant damage, the tensile energy, and the packing fraction, along with a comparison with snapshots of a matching experiment at comparable moments in time. The distribution of $d$ closely matches the fractured portion of the particle raft in the experiment. The tensile energy field $W^+$, which is representative of the processes that underpin crack propagation, transitions from an initially circular tensile distribution due to the surfactant force to one with broken symmetry depending on the available energy in the system. The tensile energy drives the three-branches until they eventually arrest due to an increase in the compressive energy. The field $\phi$ depicts the evolution of the initial packing fraction. As the fractured region expands, the monolayer contracts (to approximately $\phi=0.87$) and the mobility of the surfactant in the fractured region ($\phi=0.3$) is increased.






\subsection{Phase diagram for fracture pattern characterization}
\label{sect:relation}

\begin{figure}[b!!]
   \centering
      \begin{overpic}[width=0.48\textwidth]{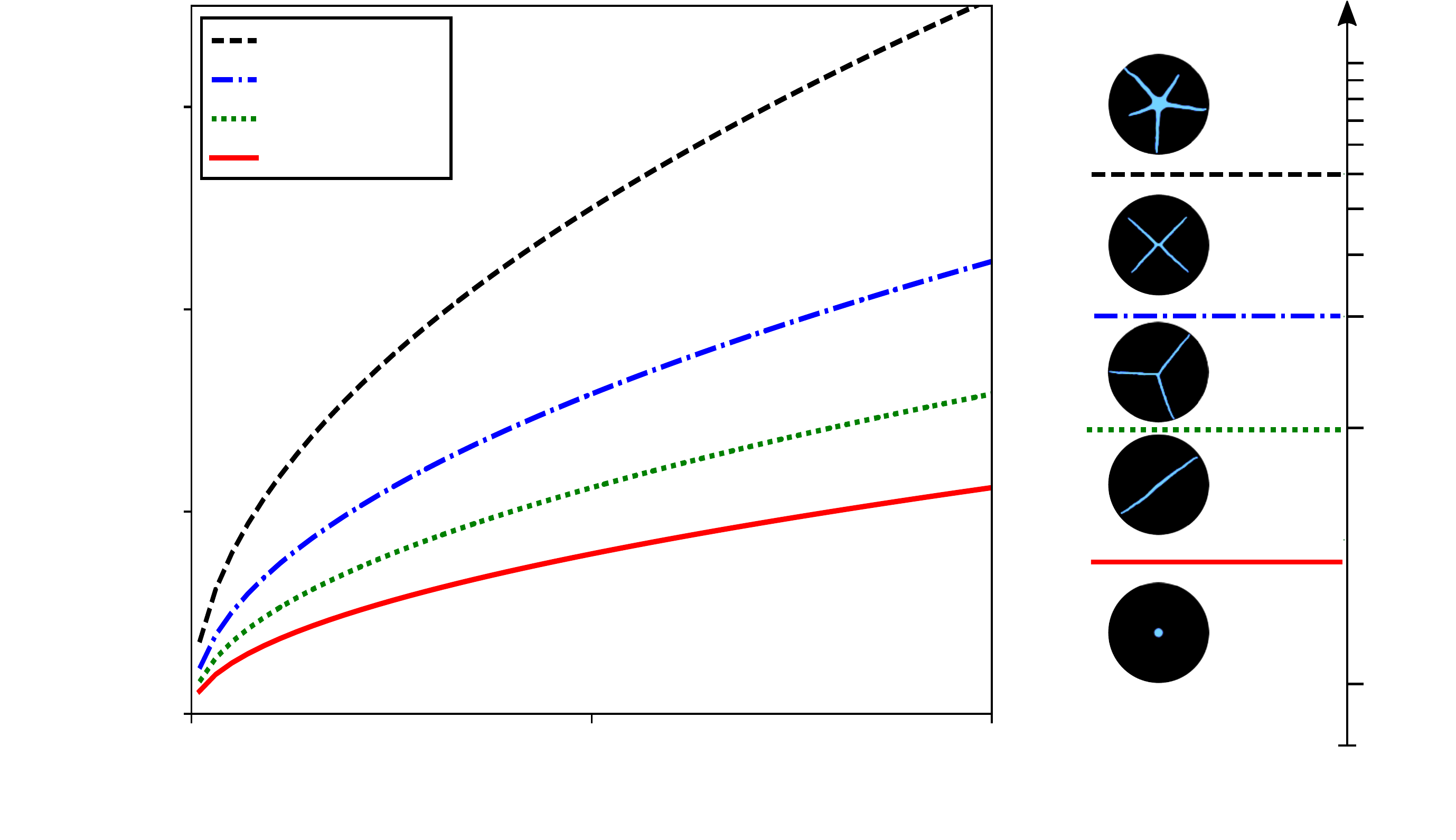}


\put (20,54.5) {\scriptsize$\chi=0.050$}
\put (20,52) {\scriptsize$\chi=0.020$}
\put (20,49.3) {\scriptsize$\chi=0.010$}
\put (20,46.5) {\scriptsize$\chi=0.005$}

\put (85.5,20) {\footnotesize$0.005$}
\put (85.5,29) {\footnotesize$0.010$}
\put (85.5,37) {\footnotesize$0.020$}
\put (85.5,46.7) {\footnotesize$0.050$}

\put (94,27.5) {\footnotesize$10^{-2}$}
\put (94,52.7) {\footnotesize$10^{-1}$}

\put (9.5,5) {$0.00$}
\put (37,5) {$0.05$}
\put (64,5) {$0.10$}

\put (5.8,21.5) {$0.02$}
\put (5.8,35.5) {$0.04$}
\put (5.8,49) {$0.06$}

\put (1,30) {$F^*$}
\put (39,1) {$G^*$}

\put (91.5,3.2) {$\chi$}


\put (50,30) {\footnotesize zone iii}
\put (40,37) {\footnotesize zone iv}
\put (59,25) {\footnotesize zone ii}

\put (44,15) {\footnotesize zone i}
\put (40,12) {\footnotesize (no fracture)}
\put (20,42) {\footnotesize zone v}
\put (15,39) {\footnotesize (multifracture)}

\end{overpic}
   \caption{Phase diagram for fracture pattern characterization. Threshold levels of the dimensionless number $\chi = (F^*)^2/G^*$  delineate various patterns. In the region below $\chi=0.005$ (zone i), the initially fractured region does not grow with time. As $\chi$ is increased, the fracture patterns gradually change from the single branch configuration (zone ii) to the three (zone iii), four (zone iv) and five (zone v) branch configurations.}
    \label{fig:phase_2}
\end{figure}
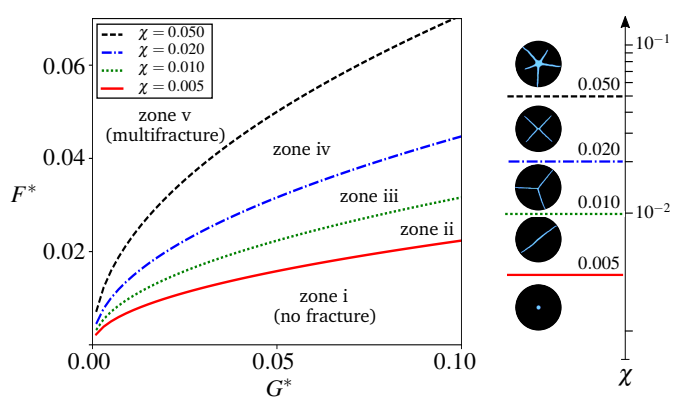

We now study whether the dimensionless parameter $\chi$ can be used to construct a phase diagram for the fracture patterns that are produced.  
In so doing, we eliminate the random aspect of the previous simulations and consider initial packing fractions that are spatially uniform with $\phi_0 = 0.80$.
As illustrated in Figure~\ref{fig:phase_2}, the results allow us to identify threshold levels of $\chi$ that delineate regions of a phase diagram for the fracture patterns obtained through simulations.
 The threshold $\chi=0.005$ defines a limit below which fracture is not observed. As $\chi$ is increased, the fracture pattern starts to develop as a single crack, and then into three-branch shapes with further increases. We note that the three-branch configuration is quite ubiquitous in nature, as the presence of acute angles and the compression generated by the three cracks stabilizes the system. For larger values of $\chi$, a fourth and a fifth branch are gradually realized. Larger values of $\chi$ are required to produce larger number of crack branches, resulting in unstable fracture patterns that cannot be clearly delineated in well defined regions of our phase diagram.

\begin{figure}[t!!]
   \centering
   \begin{overpic}[trim=3cm 0cm 3cm 0cm,height=5cm]{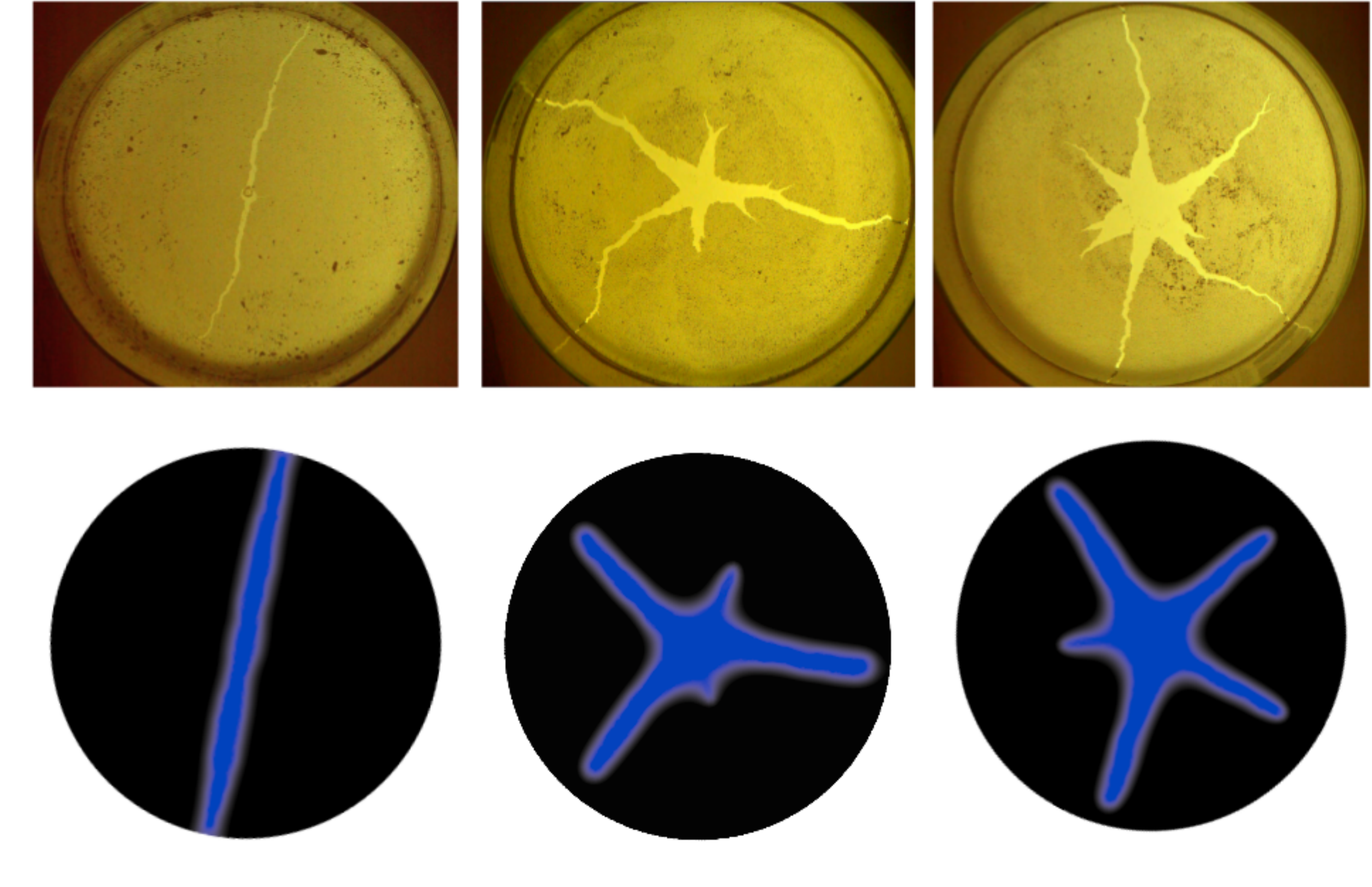}
      \put (48,-2) {$\chi_{\textit{ref}}$}
            \put (1,-2) {$\chi_{\textit{down}}$}
                  \put (92,-2) {$\chi_{\textit{up}}$}
   \end{overpic}
   \caption{Comparison between fracture patterns from experiments (top) and simulations (bottom).  In all cases, the final configuration is shown.  Simulation results were obtained for values of the dimensionless parameter $\chi$ ($\chi_{\textit{down}}=0.006$, $\chi_{\textit{ref}}=0.018$, $\chi_{\textit{up}}=0.035$), achieved by modulating the surface tension of the liquid layer and the surfactant used in each of the experiments, consistent with \eqref{eq:chi2}.   }
    \label{fig:comp2}
\end{figure}

To validate the zones and thresholds predicted by the simulations, we first calibrated our model to experiments using oleic acid spreading over a pure water layer.  The resulting fracture pattern is shown in the bottom of the center column of Figure~\ref{fig:comp2}, which presents images of the final configurations for three different experiments, along with simulation results for comparable values of $\chi$. Numerical simulations indicate that such a pattern can be obtained using $\chi = 0.018$, which we label as $\chi_{ref}$.  As indicated by expression (\ref{eq:chi2}), changing the surface tension difference corresponds to modulating the value of $\chi$ in the model.   Accordingly, we conducted two additional experiments.  In the first of these experiments, the surface tension of the liquid layer was reduced from $72$ to $56$ mN/m (colsolvent).  In the second, the  surface tension of the surfactant was reduced from $40$ to $28$ mN/m (acetone). These changes in surface tension correspond approximately to modifying factors of $1/3$ and $2$, respectively, on the reference $\chi_{\textit{ref}}=0.018$.  We denote the corresponding levels of $\chi$ as $\chi_{\textit{down}}$ and $\chi_{\textit{up}}$.




As illustrated in Figure~\ref{fig:comp2}, simulations based on the values of $\chi_{\textit{up}}$ and $\chi_{\textit{down}}$ yield fracture patterns that are remarkably consistent with the corresponding experiments.  Simulations based on doubling $\chi$ from $\chi_{\textit{ref}}$ to $\chi_{\textit{up}}$ result in an increase in the number of fracture crack branches, from three to five.  Conversely, simulations based on a reduction of $\chi$ from $\chi_{\textit{ref}}$ to $\chi_{\textit{down}}$ result in a decrease in the number of crack branches, from three to two.  In each case, a qualitative match with the corresponding experiment is clear. As $\chi$ is increased, the model presents a gradual change between patterns, from the formation of a single crack to a configuration with multiple branches. The model nicely reproduces the central damage zone and multiple crack branches observed in experiments for the range of $\chi$ in this study ($10^{-4}$--$10^{-1}$).  


This result is particularly important because the conventional view has been that the packing fraction distribution is the primary variable governing the morphological features of the crack pattern.  Our results suggest that, although the randomness of the initial packing fraction can underpin features such as crack deflections and arrest, the branching and general patterns observed in the experiments correspond to a much more fundamental principle related to the balance between the surfactant force and the fracture toughness.  In other words, the difference in surface tension between the surfactant and underlying liquid are central to determining the crack patterns that emerge.  


\subsection{Temporal scaling}
\label{sect:tubes}

\begin{figure}[tb!]
 	\vspace{-1cm}
   \centering
    \begin{overpic}[width=0.49\textwidth]{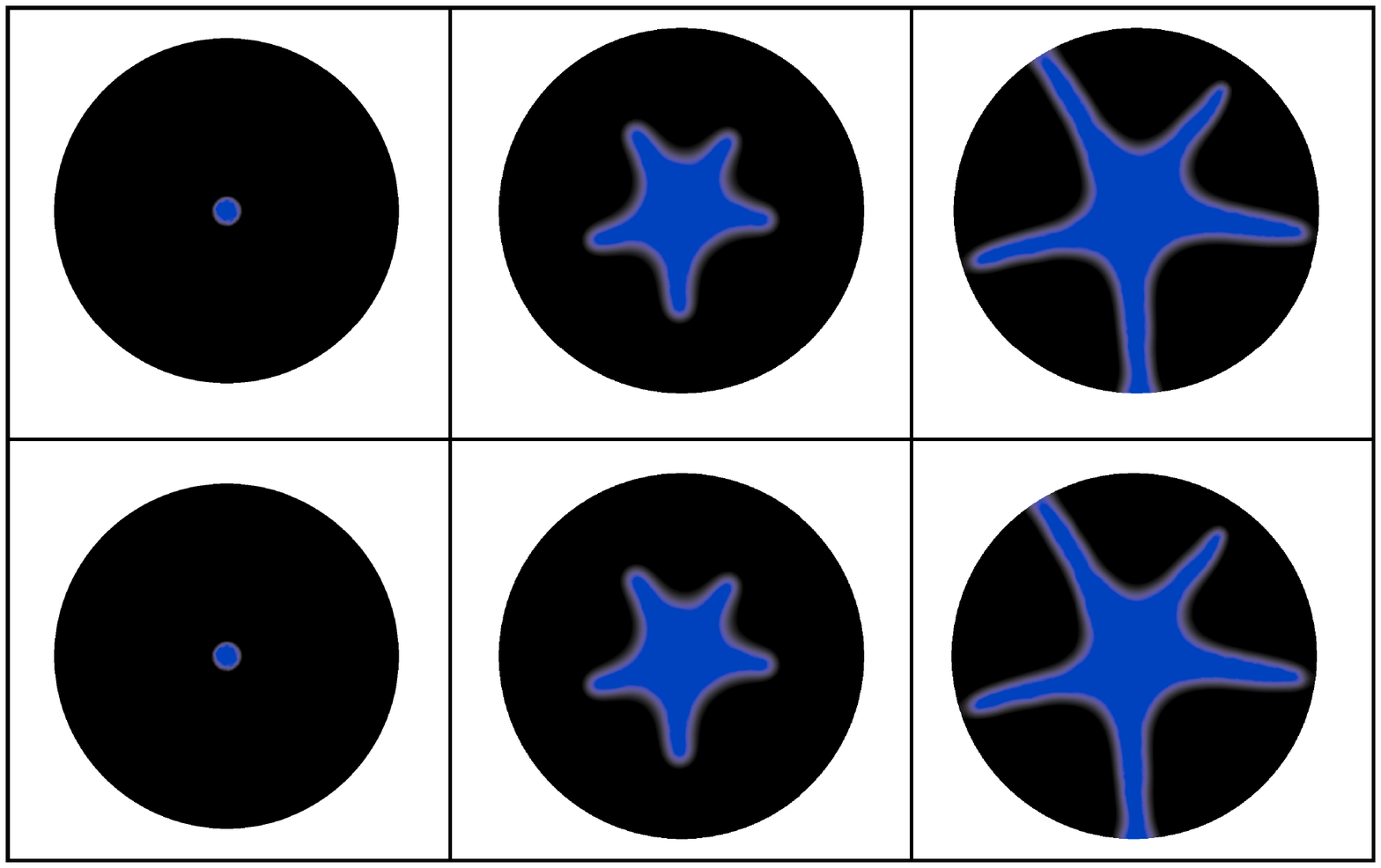}
  \put (10,45) {$\mu_1$} 
\put (10,23) {$\mu_2$} 
\put (21,57) {$t = 0.0 s$} 
\put (21,11.5) {$t = 0.0 s$} 

 \put (43,57) {$t = 0.5 s$} 
\put (43,11.6) {$t = 50 s$} 
\put (64,57) {$t = 1.00 s$} 
\put (64,11.5) {$t = 100 s$}          
\end{overpic}\vspace{-1.0cm}
   \caption{Identical fracture patterns are produced at different speeds due to changes in the viscosity $\mu_f$ of the underlying liquid layer.}
    \label{fig:comp5}
\end{figure}

We now investigate the extent to which the model captures the speed of crack propagation.  We examine the overall time required to fracture the particulate monolayer as well as the temporal scaling of the process.   The snapshots in Figure~\ref{fig:comp5} show two fracture patterns that evolve at very different rates, but for which the final configurations are visually indistinguishable. For comparison purposes, we show our results using a dimensional version of the model (i.e.,~viscosity in Pa s). Previous experimental observations have indicated that, for the same initial packing fractions, the geometry of the fracture pattern is insensitive to the viscosity of the underlying liquid layer.\cite{Vella2006} In the experiments, the liquid viscosity can be adjusted, for example, by increasing the ratio of glycerol in the glycerol-water mix based liquid layer,  raising the viscosity from $10^{-3}$ Pa s to $10^{-1}$ Pa s. This is consistent with work showing that the rate at which the process occurs decreases by the same order of magnitude.\cite{Vella2006} We can capture this effect by adjusting the mobility defined in (\ref{eq:mobility}), which is inversely proportional to the viscosity $\mu_f$ of the liquid layer.  
 
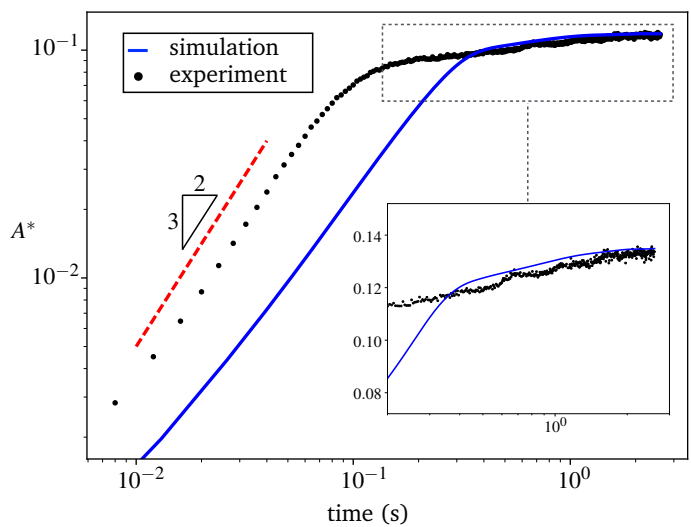
\begin{figure}[tb!]
 	\centering
 	\begin{overpic}[width=0.47\textwidth]{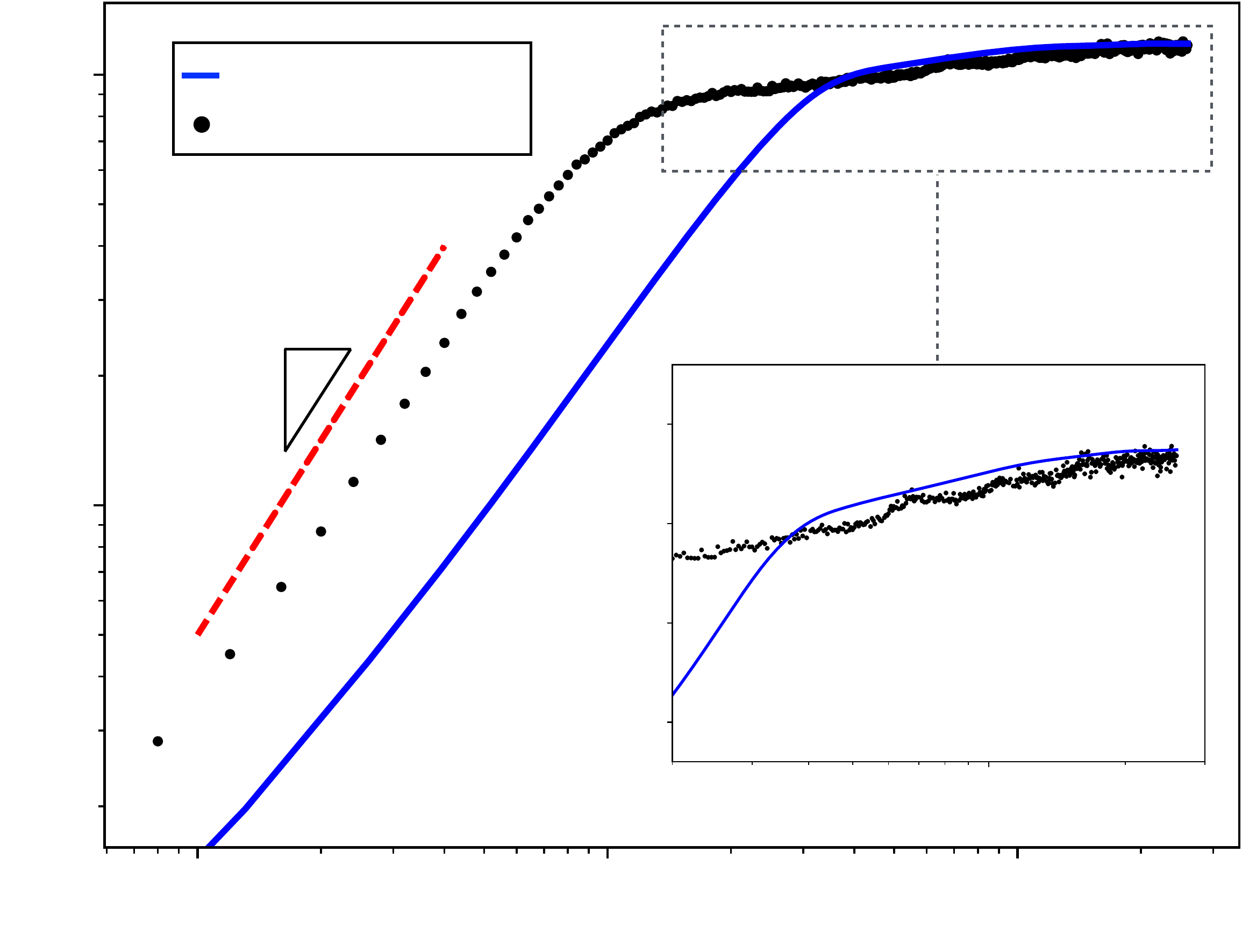}
 		\put (21,70) {simulation}
 		\put (21,65.8) {experiment}

 		\put (45,-1) {time (s)}
 		
 		\put (0,70) {$10^{-1}$}
 		\put (0,35) {$10^{-2}$}
 		
 		\put (13,4) {$10^{-2}$}
 		\put (47,4) {$10^{-1}$}
 		\put (80,4) {$10^{0}$}
 		
 		\put (24.7,49.2) {$2$}
 		\put (20.7,43.5) {$3$}
		
			\put (78,12.5) {\scriptsize $10^{0}$}
			\put (49,17.5) {\scriptsize $0.08$}
			\put (49,25.5) {\scriptsize $0.10$}
                         \put (49,33.5) {\scriptsize $0.12$}
                        \put (49,41.5) {\scriptsize $0.14$}

 		\put (-3,42) {$A^*$}      
 	\end{overpic}        
   \caption{Comparison between simulation and experiment  for the temporal scaling for the normalized fractured area $A^*$ (the ratio of the portion of the area of the liquid layer that is exposed to the area of $\Omega$), indicating two regimes of crack evolution.  For reference, a $t^{3/2}$ curve is provided (red dashed line). The inset provides a zoom of the latter stages using semi-log axes, revealing a slow logarithmic relaxation of the fracture pattern.}
    \label{fig:comp4}
\end{figure}

Additionally, experimental measurements indicate that the fracture proceeds in two stages.  Crack growth rates that scale with the $3/4$ and higher powers of $t$ are observed in the early stages of the fracture process, followed by a gradual flattening as the crack advances and eventually arrests.\cite{Vella2006}  Arrest may occur in response to the increase in packing around the perimeter of the domain,  which gradually makes it more difficult to displace the particles of the monolayer.
 As shown in Figure~\ref{fig:comp4}, our model qualitatively reproduces this behavior with time, in terms of the normalized fractured area $A^*$ (namely the ratio of the area of the surface of the liquid layer that is exposed to the area of $\Omega$). As argued by Bandi et al.,\cite{bandi2011} if the radially receding front of $R(t)$ of a particle raft scales according to $R(t)\sim t^{3/4}$, then the particle free area proceeds as $R(t)^2\sim(t^{3/4})^2\sim t^{3/2}$.  The growth rates in the early stages are slightly below $t^{3/2}$, which suggests the presence of the monolayer slightly retards the expansion of the surfactant from what would be expected over a pure liquid surface.\cite{Jensen1995}  Finally, we recognize a shift between the results from experiment and simulation with the  initial expansion process being noticeably faster in the experiment.  This discrepancy at early times could be related to the rapid expansion in the drop of surfactant that is not captured by the model.  The subtleties of the physics at early times are likely sensitive to the volume and height of the drop of surfactant at the time of injection.  Questions related to these subtleties are a subject for future work.

Following the rapid initial expansion, the dynamics slow down considerably once the particulate rafts jam together. As shown in the inset to Figure~\ref{fig:comp4}, the late stage dynamics exhibit approximately logarithmically slow relaxation over a decade in time. This kind of dynamics, also known as ``creep'' or ``aging'', is observed in a variety of phenomena including granular compaction,\cite{Ben-Naim1998} electron glasses,\cite{Amir2009} polymer relaxation,\cite{ferry1980} and superconductors,\cite{Gurevich1993} and is now considered a hallmark of non-exponential, slow relaxation in amorphous media.\cite{Amir2012} Although the observation times in our model and experiments do not extend over several decades, they are consistent with the expected behavior of slow relaxation in frustrated, amorphous media.

 

 
\section{Conclusions}
\label{sect:conc}

We have presented a phase-field model that captures the main features of the surfactant-driven fracture of closely-packed particulate monolayers. 
The model and experimental results indicate that there is a competition between the spreading force of the surfactant and the fracture toughness of the monolayer that determines the number of fractures.  The model gives rise to a dimensionless parameter that can be written in terms of the surface tensions, and which allows for a straightforward comparison with experimental conditions.  Experiments were conducted to validate the model and delineate the regimes of fracture pattern as a function of that parameter.  
  
 Our model rests on a number of simplifying assumptions.  In particular, we used the damage to the particulate monolayer as an indicator function for the surfactant concentration.  As such, our model precludes investigations of the extent to which the surfactant can penetrate into the particulate monolayer network, ahead of (or behind) the crack front.  We have also assumed that the strains exhibited by the monolayer are sufficiently small to justify modeling it as a linearly elastic solid.  This assumption ceases to be reasonable if the surface tension difference is increased above some critical threshold.   Moreover, its applicability is limited to systems in which the initial packing is low enough to allow the particulate layer to exhibit both loosely-packed and jammed behaviors as the surfactant spreads.\cite{bandi2011} In either case, transitions between linear and nonlinear response, or between fluid and solid behavior, are worthy of further study.  
Finally, the interplay between liquid and solid constituents in the monolayer may endow it with a viscoelastic behavior that could also be explored.  As the rate of surfactant transport is a function of the Marangoni stress, modulating the surface tension difference might also be used to explore any rate dependence in the monolayer.  Future work will focus on enhancing the model to incorporate these effects and to allow for detailed exploration of their consequences and significance.  
 
 
 
 
 

\section{Acknowledgments}

Christian Peco, Yingjie Liu, and John E.\ Dolbow are grateful for the support of NSF grant CMMI-1537306, to Duke University.  Wei Chen, M. M. Bandi, and Eliot Fried gratefully acknowledge support from the Okinawa Institute of Science and Technology Graduate University with subsidy funding from the Cabinet Office, Government of Japan.  John E.\ Dolbow would also like to acknowledge support from the Okinawa Institute of Science and Technology during a sabbatical stay in late 2014, during which time this work began.




\balance


\bibliography{praft_fracture} 
\bibliographystyle{rsc} 

\end{document}